\documentclass{aa}  

\usepackage{graphicx}
\usepackage{txfonts}

\begin{document}

   \title{Evolution of the kinematic properties of rotating multiple-population globular clusters}

   \author{Ethan B. White
          \inst{1},
          Enrico Vesperini
          \inst{1},
          Emanuele Dalessandro
          \inst{2} and
          Anna Lisa Varri
          \inst{3,4}
          }
    
   \institute{$^{1}$Department of Astronomy, Indiana University, Bloomington, Swain West, 727 E. 3rd Street, IN 47405, USA\\
   $^{2}$INAF - Astrophysics and Space Science Observatory Bologna, Via Gobetti 93/3, 40129 Bologna, Italy\\
   $^{3}$Institute for Astronomy, University of Edinburgh, Royal Observatory, Blackford Hill, Edinburgh EH9 3HJ, UK\\
   $^{4}$School of Mathematics and Maxwell Institute for Mathematical Sciences, University of Edinburgh, University of Edinburgh, Kings Buildings, Edinburgh EH9 3FD, UK
             }

   \date{}

  \abstract
  {Globular clusters host multiple stellar populations differing in their chemical and dynamical properties. 
  A number of models for the formation of multiple populations predict that the subsystem of second generation (SG) stars (those with anomalous chemical abundances) is characterised by a more centrally concentrated spatial distribution and a more rapid rotation than the system of first generation (FG) stars (those with chemical properties similar to field stars). In this paper, we present the results of a suite of N-body simulations aimed at exploring the long-term dynamical evolution of rotating multiple-population globular clusters. We study the evolution of systems starting with four different orientations of the cluster's total internal angular momentum vector relative to the orbital angular momentum. This allows us not only to explore the internal evolution driven by two-body relaxation, but also the effects of the cluster's interaction with the galactic tidal field and how this interaction affects the cluster's internal rotation over time. 
  We focus our attention on the kinematic differences between the two generations and we quantify these differences by exploring the FG and SG rotation velocity and angular momenta. We find that kinematic differences between the generations persist for a majority of the simulations' lifetimes, although the strength of these differences rapidly decreases after a few relaxation times. The differences can be seen most clearly in the lowest-mass stars in the models.
  We find that the clusters' internal angular momentum gradually aligns with the orbital angular momentum over time, although there is little difference in this alignment between the FG and SG systems. We also find that stars in the cluster's outer regions align with the orbital angular momentum vector more rapidly than those in the inner regions leading to a variation of the orientation of the internal angular momentum with the clustercentric distance. The alignment between internal angular momentum and orbital angular momentum occurs more rapidly for low-mass stars. We also study the evolution of the anisotropy in the velocity distribution and, in agreement with previous results, find the SG to be characterised by a stronger radial anisotropy than the FG. Overall, our results show that the kinematic properties of multiple populations provide key information on their formation and dynamical evolution.}

   \keywords{Globular clusters: general - Stars: kinematics and dynamics}
\titlerunning{Evolution of rotating globular clusters}
\authorrunning{E.White et al.}
   \maketitle
%

\section{Introduction}

Numerous observational studies in the last few decades have provided new insights into the stellar content of globular clusters and revealed that these systems host multiple stellar populations differing in the abundance of several light elements (such as Na, O, Al, Mg, C, N, and He) (see e.g. \citealt{carretta2009a, carretta2009b, marino19, carretta19}; see also \citealt{gratton12} and references therein). About 20 percent of Galactic clusters have been found to be even more complex and also display a spread in Fe (see e.g. \citealt{milone17}). 

Despite the extensive investigations and progress made in this field, many questions concerning the origin of multiple populations and the formation history of globular clusters are still unanswered. 

The various formation pathways proposed in the literature consider different sources of processed gas for the formation of the chemically anomalous population including AGB stars, massive binary stars, rapidly rotating stars, and supermassive stars (see e.g. \citealt{mink09, dercole10, Krause13, denissenkov13, wang20, nguyen24}). Models differ also in the star formation history of multiple populations with some scenarios proposing separate star formation events (see e.g. \citealt{D_Ercole_2008, mink09, denissenkov13, d'antona16}) and others in which the different populations are instead formed at the same time (see e.g. \citealt{bastian13, gieles18}). Hereafter we will refer to chemically anomalous stars formed from processed gas as second-generation (SG) stars and to stars with chemical composition similar to that of halo field stars as first-generation (FG) stars. 

Despite these differences, most models predict that SG stars would form centrally concentrated in the inner regions of a more spatially extended FG system (see e.g. \citealt{D_Ercole_2008, calura19}). A few studies have investigated the SG formation in a FG cluster characterised by internal rotation and have shown that in this case the two populations differ also in their kinematic properties with the SG system characterised by a more rapid rotation than the FG system (see e.g. \citealt{bekki10}, \citeyear{bekki11}; \citealt{lacchin22}). 

In order to link the present-day dynamical properties of FG and SG stars with those imprinted at the time of their formation, it is necessary to study how various early- and long-term evolutionary processes affect the structure and kinematics of globular clusters and their multiple populations. 

A number of studies have been focused on the theoretical investigation of the evolution of the initial dynamical differences of FG and SG subsystems, their dependence on the initial conditions and the role played by various dynamical processes (see e.g. \citealt{vesperini13, mastrobuono13}, \citeyear{mastrobuono16}; \citealt{henault15, vesperini18, Maria, livernois24, hypki24, berczik24, aros25}). On the observational side, many studies have detected spatial and kinematic differences in the present-day properties of several Galactic globular clusters (see e.g. \citealt{Bellini_2015, milone18, dalessandro19}, \citeyear{dalessandro21}, \citeyear{dalessandro24}; \citealt{libralato23, leitinger23, martens23, cordoni25, ziliotto25, griggio25}) and revealed trends between the extent of these differences and the clusters' dynamical ages (see e.g. \citealt{dalessandro18}, \citeyear{dalessandro19}, \citeyear{dalessandro24}; \citealt{libralato23, cordoni25}). 

The goal of this paper is to investigate the long-term dynamical evolution of multiple generations in globular clusters with internal rotation through the use of N-body simulations. Specifically, we seek to explore the spatial and kinematic properties of these multiple generations, how they depend on stellar mass, the role of the initial orientation of the internal rotation axis relative to the cluster's orbital angular momentum, and what this can reveal about the cluster's initial conditions. 

The outline of this paper is as follows. In Section 2, we will describe the methods and initial conditions of our simulations; in Section 3, we will describe the results of our analysis; finally, in Section 4, we summarise the conclusions drawn from our analysis.

\section{Methods and initial conditions}

   We performed a set of N-body simulations using \texttt{NBODY6++GPU} (\citealt{Wang_2015}) on the Big Red II and Big Red 200 supercomputers at Indiana University. 

The models in our simulations move on circular orbits around the centre of their host galaxies and the potential of the host galaxy is modelled as that of a point mass. 

Each of our models begins with 100,000 stars with masses between 0.1 and 1 \(M_\odot\) distributed using a \cite{kroupa2001} initial mass function. Our investigation is focused only on the long-term evolution driven by two-body relaxation and the effects of the host galaxy tidal field; effects due to mass loss associated with stellar evolution are not included and, given the mass range included in our models, would be expected to be low. Our N-body models begin with an equal number of FG and SG stars; in a number of scenarios for the formation of multiple stellar populations the initial number of SG stars is smaller than the FG but the ratio of the SG to FG number rapidly increases to values close to the one as adopted here or larger during the cluster's early evolution (see e.g. \citealt{D_Ercole_2008, vesperini21, sollima21, hypki24}).

The initial structural properties of our models are characterised by a rapidly rotating, flattened and centrally concentrated SG subsystem embedded in a more diffuse, slowly rotating FG system. These structural properties are based on the results of a number of studies based on hydro/N-body simulations following the formation of multiple stellar populations in rotating clusters (\citealt{bekki10}, \citeyear{bekki11}; \citealt{lacchin22}).

Our initial conditions have been set using the code \texttt{\textsc{MaGalie}} (\citealt{BOILY200127}): the SG subsystem is modelled as a rotating exponential disc while the FG system follows a spherical distribution with a Hernquist density profile (\citealt{hernquist93}). Rotation is included in the FG system by following the Lynden-Bell demon procedure (\citealt{lynden_bell}): we reverse the rotation velocity around the rotation axis of the cluster of 50$\%$ of retrograde stars in the FG such that 75$\%$ of stars in the FG rotate on prograde orbits. 

The FG system extends to the cluster's tidal radius and has a ratio of the 3D half-mass radius to the tidal radius equal to about 0.085. The SG subsystem is initially centrally concentrated and the ratio of the FG to the SG maximum (half-mass) radii is initially equal to 5 (5.93). Stars are removed from a simulation once they pass beyond a radius equal to twice the cluster's Jacobi radius.

In order to explore the effects of the external tidal field and its torque on the internal rotational kinematics of the multiple stellar populations, we have run a set of simulations with different orientations of the cluster's initial total internal angular momentum vector relative to the orbital angular momentum vector associated to the cluster's orbit around the host galaxy (see \citealt{Maria}, \citeyear{maria22} for the first studies showing the complex kinematic features that may emerge from such a misalignment). Throughout this work, we utilise two parameters when describing the internal angular momentum. These are the internal angular momentum, which describes the angular momentum of individual stars in the cluster, and the total internal angular momentum, which describes the combined internal angular momentum of a group of stars in a specified region.

To explore the full range of dynamic possibilities resulting from this misalignment, we have run four different simulations corresponding to four different initial angles between the internal and orbital angular momentum vectors equal to 0, 45, 90, and 180 degrees (hereafter we refer to those models as Delta0, Delta45, Delta90, and Delta180; see Table \ref{tab:models}).

\begin{table}
    \caption{Models}
    \centering
    \begin{tabular}{||c c||}
         \hline
         Model ID & $\delta$ (degrees) \\
         \hline
         Delta0 & 0  \\ 
         Delta45 & 45 \\
         Delta90 & 90 \\
         Delta180 & 180 \\
         \hline
    \end{tabular}
    \tablefoot{The names for the N-body simulations used in this study. The symbol $\delta$ denotes the initial angle between the model's total internal and orbital angular momentum vectors. $\delta$ is measured from the orbital angular momentum vector.}
    \label{tab:models}
\end{table}

All the kinematic quantities presented in this study have been calculated in an inertial non-rotating reference frame. At any given time, all dynamical quantities have been calculated by combining five sequential snapshots of a system at times centred around the chosen time to improve statistics of the calculated quantities. The only exception to this is when looking at quantities calculated at a time of t = 0, where the initial conditions are used without combining multiple snapshots.  

Our analysis of a cluster's spatial and kinematic properties is performed in a cylindrical coordinate system defined as follows: the z-axis of this system is aligned with the direction of the total internal angular momentum vector of stars within the 2D half-mass radius of the aggregate snapshot; the other two coordinates are the standard radial and angular coordinates on the plane perpendicular to the z-axis with the radius on this plane corresponding to the standard projected distance from the cluster's centre; radial and tangential velocities used in the rest of the paper refer to the velocities in these two components. In particular, for the analysis of the internal angular momentum distribution and evolution, we calculate the rotation velocity profiles using the tangential velocities on the plane defined above. 
This calculation corresponds to the observational determination of rotation velocity using proper motion measurements adopting an ideal line of sight parallel to the cluster's total internal angular momentum. The calculation of the total internal angular momentum for this purpose includes only stars within the cluster's 2D half-mass radius; this is done to avoid the possible effect of stars in the outer regions of the system where the torque of the external tidal field may cause the internal angular momentum to be orientated differently from that of the stars in the inner regions (see \citealt{Maria}, \citeyear{maria22} for a discussion of this effect).

For the analysis of the time evolution of the kinematic properties, time is normalised to the half-mass relaxation timescale (see e.g. \citealt{heggie03}) defined as:

\begin{equation}
    t_{\rm rh} = \frac{0.138M^{1/2}r_{\rm h}^{3/2}}{ln(0.02N)G^{1/2}<m>}.
	\label{eq:thr}
\end{equation}

where N is the number of stars in the system, $M$ is the mass of the system, $<m>$ is the mean stellar mass, $r_{\rm h}$ is the 3D half-mass radius, and G is the gravitational constant. 
 
In some of our analysis, we fit the rotation curve radial profiles (see Sections 3.2 $\&$ and 3.3) using the following analytical expression (\citealt{lynden67}).

\begin{equation}
    V_{\rm rot} = \frac{2Ra_{\rm rot}}{R_{\rm peak}[1 + (\frac{R}{R_{\rm peak}})^2]}
    \label{eq:rot}
\end{equation}
Here $V_{\rm rot}$ is the rotation velocity, R is the projected distance from the centre of the cluster, $R_{\rm peak}$ is the location of the peak velocity, and $a_{\rm rot}$ is the peak rotation velocity.

\begin{figure}
	\includegraphics[width=\columnwidth]{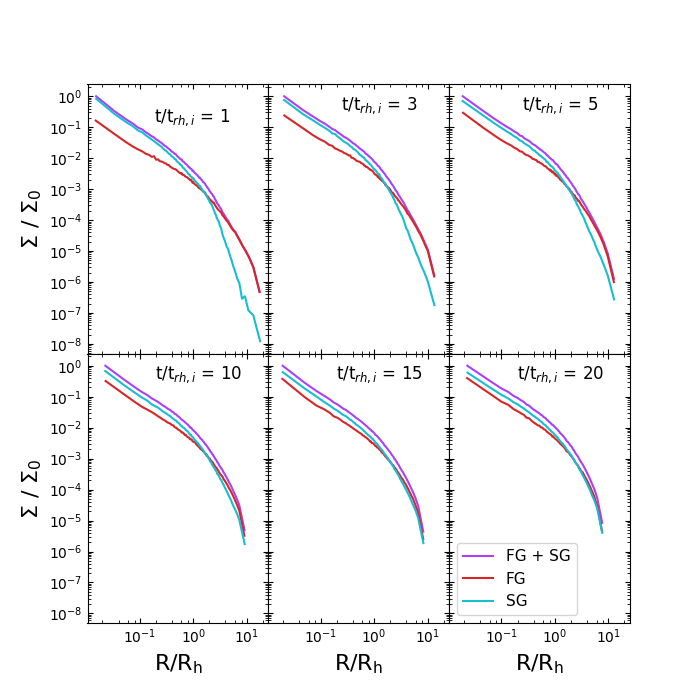}
    \caption{Surface mass density profiles as a function of the projected distance from the cluster's centre normalised by the projected half-mass radius (R${\rm_h}$) of the total cluster (see Methods for a complete definition). The profiles are shown at a few representative times in the Delta0 model. All densities are normalised by the central bin of the total surface mass density (FG + SG). Time is normalised by the initial half-mass relaxation time of the cluster, $t_{\rm rh,i}$. The difference between the profiles of the two generations is gradually erased as the system evolves.}
    \label{fig:masden_prof}
\end{figure}

\begin{figure}
	\includegraphics[width=\columnwidth]{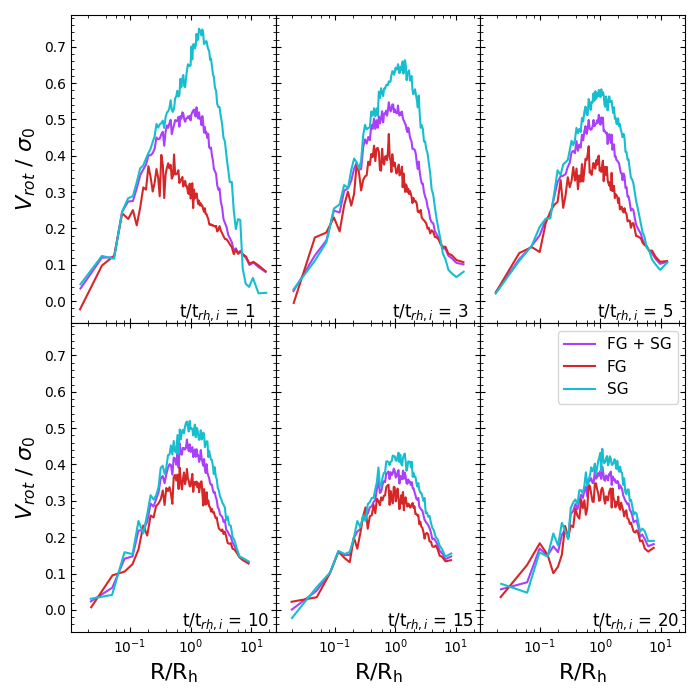}
    \caption{Radial profiles of the rotation velocity are shown at representative times for the Delta0 model. The rotation velocity is normalised by the central velocity dispersion (inner 1$\%$ of stars) of all stars (FG + SG). The radius is normalised by the projected half-mass radius (R$_{\rm h}$) of all stars (FG +SG). Time is normalised by the initial half-mass relaxation time of the cluster, $t_{\rm rh,i}$.}
    \label{fig:rot_prof}
\end{figure}

\section{Results}
\subsection{Evolution of internal kinematics}

We begin our analysis by exploring the simplest case, in which the total internal angular momentum vector of stars within the cluster's Jacobi radius and the orbital angular momentum vector are initially aligned (referred to as the Delta0 model).
It is important to note, however, that the results presented in this section are generally similar for all models we have studied. 

The two populations have initially different spatial and kinematic properties as dictated by the initial conditions. Fig. \ref{fig:masden_prof} shows how the mass density of the different generations evolves through the effects of two-body relaxation and the spatial characteristics become more similar over time. 

Fig. \ref{fig:rot_prof} shows the radial profile of the rotational velocity. This figure shows that the rotation velocity decreases as a result of two-body relaxation as the system evolves and stars are ejected from the cluster, carrying away angular momentum. This is in agreement with previous results (\citealt{einsel}; \citealt{ernst}; \citealt{kim}; \citealt{maria17}). As the system evolves and the overall rotation velocity decreases, the rotation velocities of the two populations also become more similar. We point out that while the rotation curves of the two populations still differ after several relaxation times, the strength of this difference rapidly decreases, suggesting that even relatively small present-day differences would be indicative of much stronger primordial differences imprinted at the time of the cluster's formation. We will further discuss this point later in Section 3.2 (see also the discussion of \citealt{dalessandro24}). It is interesting to point out that \cite{berczik24} also find that, in N-body models evolving on circular orbits similar to those used in this work, memory of the initial rotation may still be found in present-day clusters while initial differences between the rotational velocities of multiple populations are erased more rapidly for clusters on eccentric orbits.

As the system evolves, the initial differences between the spatial and rotational properties of the two populations are gradually erased. However, the extent of the differences between the spatial and rotational profiles in the later stages of evolution differ. If, for example, we consider only stars within the inner 2 half-mass radii (similar to what is typically available in observational data), the spatial properties are seen to mix much earlier than the rotational properties. For example, by approximately 10 $t_{\rm rh,i}$, the spatial properties in Fig. \ref{fig:masden_prof} are well mixed within the inner 2 half-mass radii, while the rotation profiles of the two generations in Fig. \ref{fig:rot_prof} are still distinct (see \citealt{dalessandro19} for an observational analysis showing the loss of spatial differences between multiple populations over time).

It is important to point out that the actual observational detection of differences between the rotational properties of FG and SG stars may be complicated by a variety of factors including the specific kinematic dataset available (line-of-sight and/or proper motion velocities), the radial range covered by the data, and the angle between the line-of-sight and the cluster's rotation axis (see \citealt{Maria} for a discussion of these points).

\begin{figure}
	\includegraphics[width=\columnwidth]{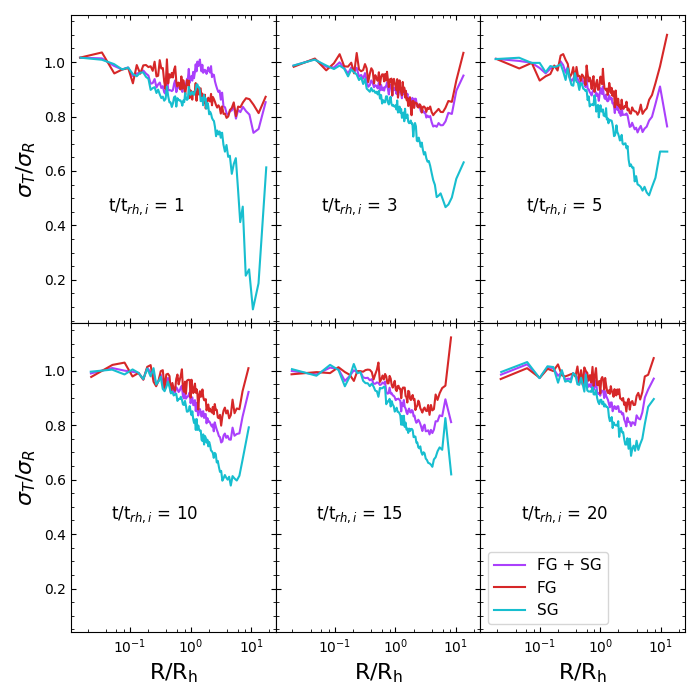}
    \caption{Radial profiles of the velocity anisotropy are shown at representative times in the Delta0 model. We define anisotropy to be the ratio of tangential velocity dispersion over the radial velocity dispersion, $\sigma_{\rm T} / \sigma_{\rm R}$, with a completely isotropic system having a value of $\sigma_{\rm T} / \sigma_{\rm R}$ = 1. Time and radius normalisations are the same as described in Fig. \ref{fig:rot_prof}.}
    \label{fig:anis_prof}
\end{figure}

In addition to the rotation velocity, we also assess the degree of anisotropy in the velocity space of the models, and its time evolution.
Fig. \ref{fig:anis_prof} shows the anisotropy profiles at the same times as shown in Figures \ref{fig:masden_prof} and \ref{fig:rot_prof}. To quantify the velocity anisotropy we use the ratio of tangential velocity dispersion to the radial velocity dispersion, $\sigma_{\rm T}/\sigma_{\rm R}$, where the tangential and the radial velocity dispersions are calculated using cylindrical bins from velocities on the plane perpendicular to the cluster's rotation axis. 

As the system evolves, the outward migration of stars on more radial orbits causes the development of a significant radial anisotropy in the cluster's outer regions; this is due to the progressive formation of a core-halo structure in collisional stellar systems due to two-body relaxation (see e.g. \citealt{heggie03}). While, for the initial conditions we have adopted, both populations develop radial anisotropy in the outer regions, the SG, which is initially more centrally concentrated, is characterised by a stronger radial anisotropy (see e.g. \citealt{Bellini_2015, Maria, vesperini21, sollima21, aros25}). The preferential loss of stars on radial orbits causes the clusters to eventually become tangentially anisotropic towards their outermost regions (see e.g. \citealt{giersz, baumgardt, maria_rad16} for a similar effect in the context of single-population clusters). 

Differences between the velocity anisotropy of first- and second-generation stars, in agreement with those found in our simulations and in previous numerical investigations of multiple-population clusters (see e.g. \citealt{Bellini_2015, vesperini21, sollima21, aros25}), have also been found in several observational studies (see e.g. \citealt{Bellini_2015, libralato23, cadelano24, dalessandro24, cordoni25}).

\subsection{Dependence of kinematic properties on the stellar mass}

We continue to focus the presentation of our results on the Delta0 model since, for this aspect of our study, we also find the results to be independent of the angle between the cluster's initial rotation axis and the cluster's orbital plane.

\begin{figure}
	\includegraphics[width=\columnwidth]{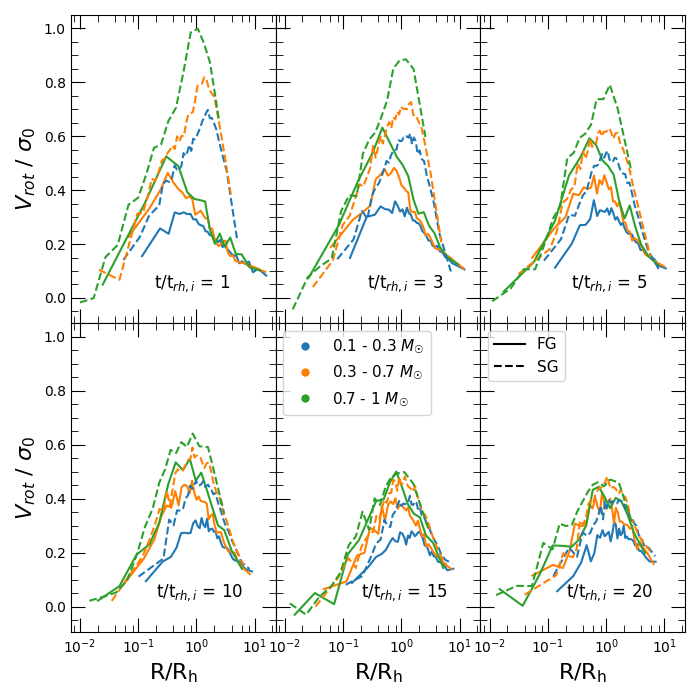}
    \caption{Radial profiles of the rotation velocity are shown at representative times for the Delta0 model for different mass groups. The line-of-sight is the same as described in Fig. \ref{fig:rot_prof}. The rotation velocity is normalised by the central velocity dispersion (inner 1$\%$ of stars) of all stars (FG + SG). Time and radius normalisations are the same as described in Fig. \ref{fig:rot_prof}. Mass groups are split into low (0.1 - 0.3 \(M_\odot\)), intermediate (0.3 - 0.7 \(M_\odot\)), and high (0.7 - 1 \(M_\odot\)) mass bins.}
    \label{fig:rot_prof_mass}
\end{figure}

Fig. \ref{fig:rot_prof_mass} shows the rotation velocity profiles for different stellar masses for both the FG and SG. Within the individual generations, mass groups initially start out with the same rotation profile; however, as the system evolves, both populations develop a relationship between rotation velocity and stellar mass where high-mass stars tend to rotate more rapidly than low-mass stars. 

While two-body relaxation causes the overall rotation to decrease for each mass bin, the lowest mass bin loses rotation velocity more efficiently. This result agrees with the recent results of \cite{Livernois_2022} who found the same behaviour in the context of single-population clusters (see also \citealt{kimleespurz}; \citealt{hong2013}; \citealt{livernois21}; see \citealt{scalco23} for the first observational evidence of this trend).

Fig. \ref{fig:rot_prof_mass} also shows that the difference between generations is erased at different rates for different mass groups. While higher-mass stars maintain a higher overall rotation velocity compared to stars with lower masses regardless of generation, the difference between the generations themselves is erased more efficiently for higher-mass stars. 
The final panel of Fig. \ref{fig:rot_prof_mass}, for example, shows that at 20 $t_{\rm rh,i}$ the rotation velocity of the two higher mass groups is nearly identical for FG and SG, while the lowest mass group is still characterised by discernible differences. 

\begin{figure}
	\includegraphics[width=\columnwidth]{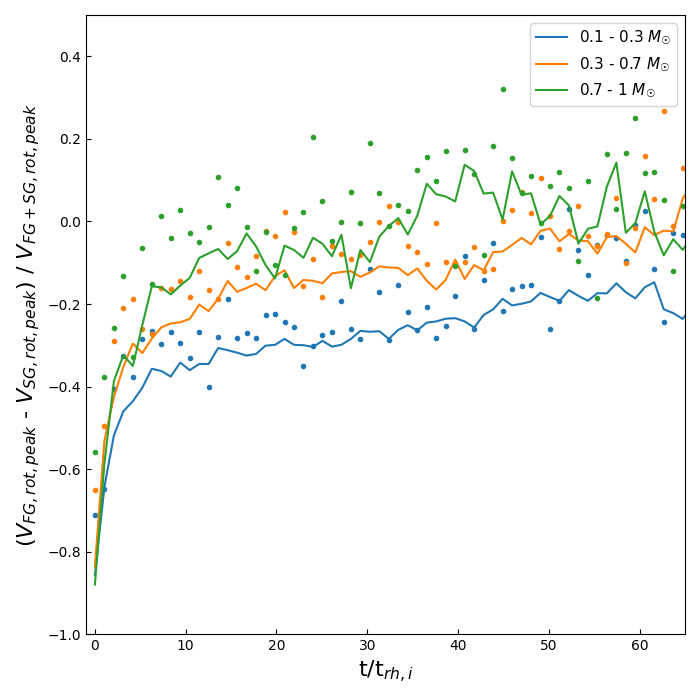}
    \caption{The time evolution of the differences between the maximum rotation velocities of the FG and SG stars is shown for different mass groups. Velocities are normalised by the peak rotation velocity for all stars (FG + SG), computed at each time. Dots correspond to the peak velocity values taken directly from the rotation profiles. Lines represent the peak velocity values taken from best-fit rotation curves to the rotation profiles (as described in Eq. \ref{eq:rot}). The time has been limited to when simulations have approximately 10$\%$ of the initial stars remaining.}
    \label{fig:rot_max_dif_mass}
\end{figure}

To further illustrate the differences between the rotational kinematics of FG and SG stars and their dependence on the stellar mass, we show, in Fig. \ref{fig:rot_max_dif_mass}, the time evolution of the difference between the FG and SG peak rotational velocities. As discussed in Section 2, we utilise best-fit lines to the rotation velocity profiles to give a clearer view of the trends. This figure clearly shows the gradual decrease of the differences between the rotation of the FG and the SG stars and how the initial differences are erased more rapidly for high-mass stars. The dependence of the strength of the rotational differences on the stellar mass implies that their detection in observational studies focused on high-mass stars (see e.g. \citealt{cordoni20a, cordoni20b, martens23, dalessandro24, leitinger25}) may not reveal significant rotational differences in clusters where such differences could still be present in low-mass stars.

We interpret this trend as the result of the interplay between the effects of mass segregation causing massive stars to migrate towards the cluster's inner regions where shorter values of the local relaxation timescale lead to a more rapid dynamical mixing of the two populations.

It is also interesting to note the rapid change in the kinematic differences at the beginning of the simulation. This suggests that caution is necessary when making claims about the initial kinematic properties of clusters based on present-day kinematics, as even in dynamically young clusters the strength of the present-day kinematic differences may be significantly smaller than the initial differences imprinted at the time of cluster formation (see also \citealt{Maria} for a discussion and cautionary remarks concerning the observational detection of differences between the rotation of the FG and the SG populations). This is true for each mass regime, even in low-mass stars where the kinematic differences persist for the longest amount of time.

\begin{figure}
	\includegraphics[width=\columnwidth]{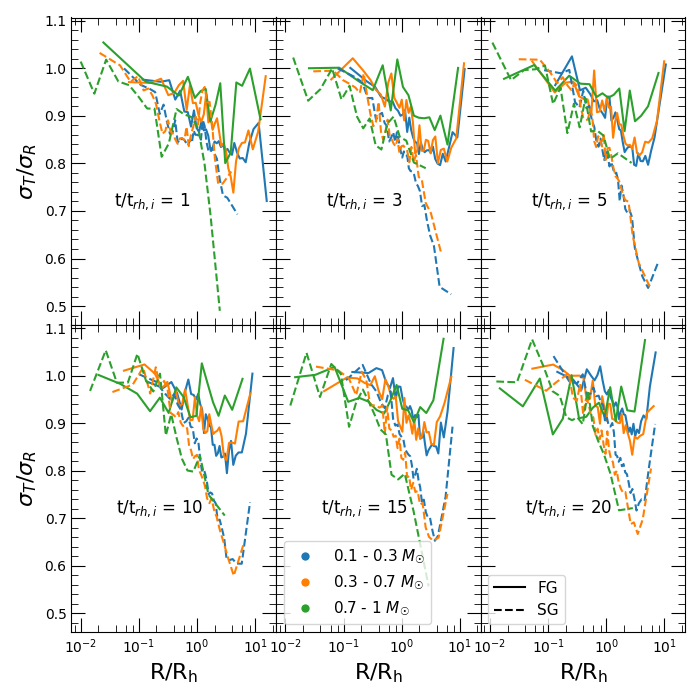}
    \caption{Radial profiles of the velocity anisotropy are shown at representative times in the Delta0 model for different mass groups. Anisotropy is defined as in Fig. \ref{fig:anis_prof}. The line-of-sight is the same as described in Fig. \ref{fig:rot_prof}. Time and radius normalisations are the same as described in Fig. \ref{fig:rot_prof}.}
    \label{fig:anis_prof_mass}
\end{figure}

Finally, in Fig. \ref{fig:anis_prof_mass} we show the radial profile of the velocity anisotropy for different mass bins. For all the mass bins, the SG is characterised by a stronger anisotropy than the FG and within each stellar population we do not find significant dependence of the degree of anisotropy on stellar mass.

\subsection{The role of the orientation of cluster's internal angular momentum relative to the orbital plane}

We continue our analysis by discussing how the initial orientation of the cluster's internal angular momentum relative to the plane of the cluster's orbit around the centre of the host galaxy affects the evolution of its internal kinematic properties. 
In particular, we explore how the differences in the initial configuration may affect the interplay between internal rotation and the torque due to the tidal field of the host galaxy. 
This can possibly lead to a more complex radial variation of the rotation pattern as the system evolves towards (partial) tidal synchronisation (see e.g. \citealt{maria18}). 

\begin{figure}
	\includegraphics[width=\columnwidth]{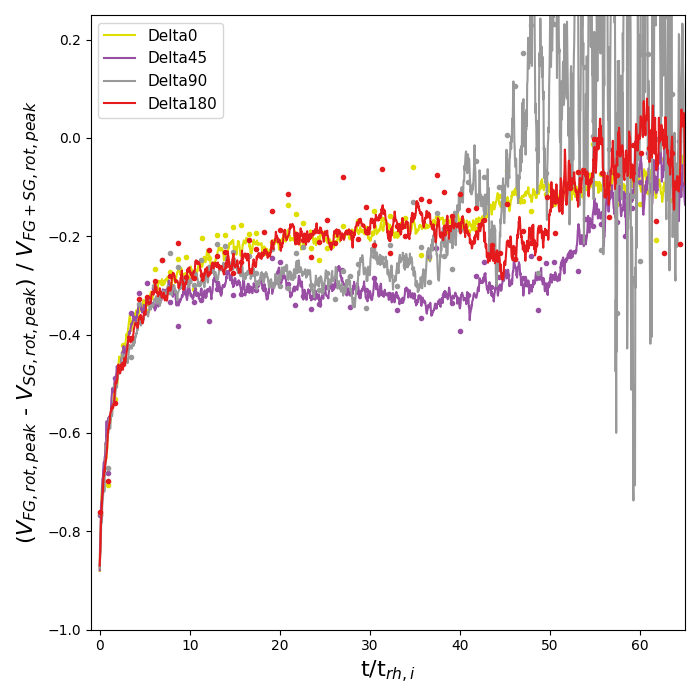}
    \caption{The time evolution of the difference between the maximum rotation velocities of the FG and SG is shown for different models including all stellar masses. Velocities are normalised by the peak rotation velocity for all stars (FG + SG). Dots correspond to the peak velocity values taken directly from the rotation profiles. Lines represent the peak velocity values taken from best-fit rotation curves 
    to the rotation profiles (as described in Eq. \ref{eq:rot}). The time has been limited to when simulations have approximately 10$\%$ of the initial stars remaining}
    \label{fig:max_rot_sims}
\end{figure}

Fig. \ref{fig:max_rot_sims} shows that the difference between the FG and SG peak rotational velocity is generally similar for the four models. However, the two models with initial rotation axis not parallel/antiparallel to the orbital angular momentum (Delta45, Delta90) are characterised by an extended phase (between about 15 $t_{\rm rh,i}$ and 40 $t_{\rm rh,i}$) during which the difference between the FG and SG rotational velocity is approximately constant or decreases at a slower rate than that of the Delta0 and Delta180 models.

\begin{figure}
	\includegraphics[width=\columnwidth]{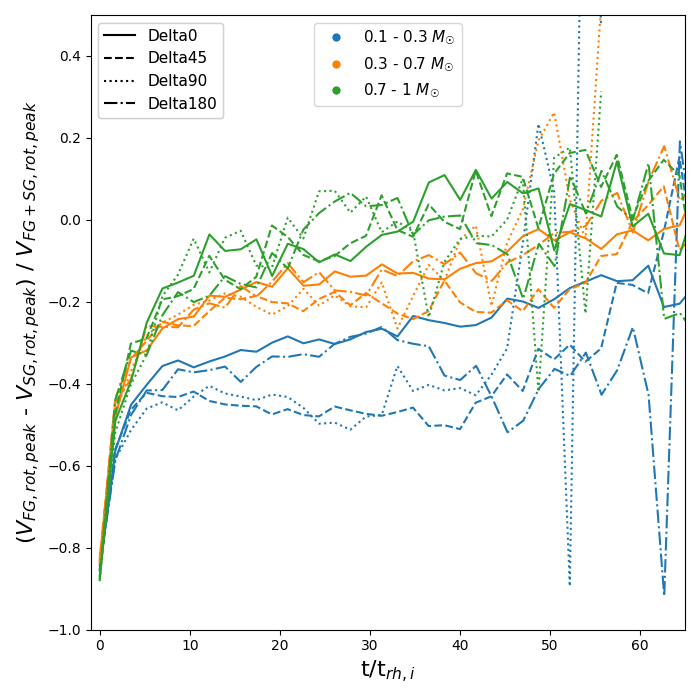}
    \caption{The time evolution of the difference between the maximum rotation velocities of FG and SG is shown for different models and relevant mass groups. The low-mass bin (0.1 - 0.3 \(M_\odot\)) is shown in blue, the intermediate-mass bin (0.3 - 0.7 \(M_\odot\)) in orange, and the high-mass bin (0.7 - 1 \(M_\odot\)) in green. Delta0 is denoted by a solid line, Delta45 by a dashed line, Delta90 by a dotted line, and Delta180 by a dash-dotted line. For the clarity of the figure, we show only best-fit lines to the data. The largest difference between generations can be seen in the lowest mass bin.}
    \label{fig:max_rot_sims_mass}
\end{figure}

Fig. \ref{fig:max_rot_sims_mass} shows the peak rotation velocity difference between generations for individual mass groups in each model. The dependence on mass is similar regardless of our choice of model. The lowest mass bin most closely resembles the behaviour seen in Fig. \ref{fig:max_rot_sims}. This is simply due to the fact that the majority of stars are those in the low-mass group and they thus have a dominant effect on the time evolution shown in Fig. \ref{fig:max_rot_sims} for the entire system.

It is important to note, however, that despite the large FG-SG difference in the low-mass stars, these are the stars which are observationally most difficult to detect. 
Most observational studies of the internal kinematics of globular clusters are instead focused on upper main sequence and red giant stars for which more accurate measurements of line-of-sight and/or proper motion velocities are possible (see e.g. \citealt{dalessandro24}). As shown by our results, the high-mass group is the one characterised by the smallest differences between the FG and SG rotation; larger differences between the rotational kinematics of multiple populations may thus be revealed by studies of the kinematics of low-mass stars.

Fig. \ref{fig:max_rot_sims_mass} also shows that all models are characterised by the rapid decrease of the initial differences between the rotational kinematics of FG and SG stars. As pointed out in section 3.2, this implies that - regardless of the initial orientation of the internal angular momentum - even in relatively dynamically young clusters the present-day differences between the FG and the SG rotational velocities are significantly smaller than the primordial ones.

\subsection{Angular momentum}

We now consider the investigation of the time evolution of the internal angular momentum of FG and SG stars and how it is affected by the external tidal field of the host galaxy. We also explore how the initial internal rotation axis relative to the orbital angular momentum affects the cluster's internal angular momentum over time. In our analysis, we explore the direction of the internal angular momentum through the use of $\delta$, the angle measured from the model's orbital angular momentum vector.

\begin{figure}
	\includegraphics[width=\columnwidth]{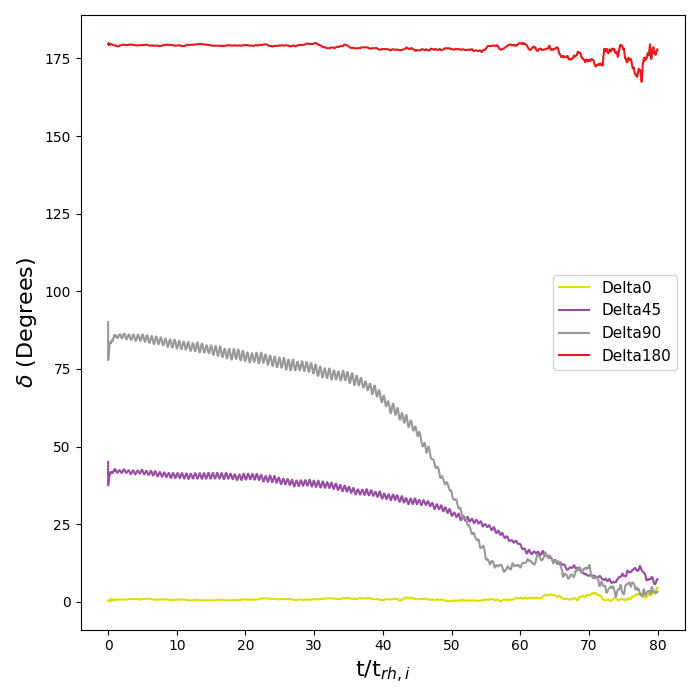}
    \caption{The time evolution of the orientation of the total internal angular momentum vector of stars within the cluster's tidal radius for different models. The total (FG + SG) internal angular momentum is shown for each model. Time is normalised by the initial half-mass relaxation time. The Delta0, Delta45, and Delta90 models each end with the orientation of their total internal angular momentum aligned largely with the orbital angular momentum at 0 degrees. The Delta180 model does not align with the orbital angular momentum by the end of the simulation.}
    \label{fig:ang_sims}
\end{figure}

\begin{figure}
	\includegraphics[width=\columnwidth]{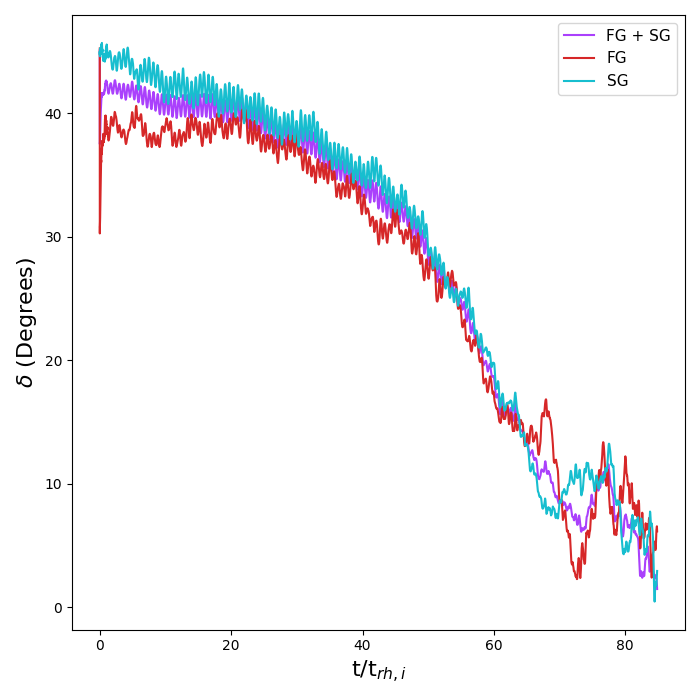}
    \caption{The time evolution of the orientation of the total internal angular momentum vectors of different subsystems including stars within the cluster's tidal radius of the Delta45 model. Time is normalised by the initial half-mass relaxation time.}
    \label{fig:Theta45_angles}
\end{figure}
Fig. \ref{fig:ang_sims} shows the time evolution of the $\delta$ angle for the total internal angular momentum of stars within the tidal radius for each simulation. 
The internal rotation of the Delta0 model begins aligned with the model’s orbital angular momentum vector and, therefore, the orientation of its internal rotation remains approximately constant during the entire simulation. The internal rotation of the Delta180 model is counter-aligned with the orbital rotation vector and also does not vary significantly for much of its lifetime. The Delta45 and Delta90 models, however, have a large change in their orientation over time due to interactions with the tidal field. The tidal field acts to align 
the internal angular momentum with the orbital angular momentum (see also \citealt{maria18}, \citeyear{Maria} for studies of the evolution of the orientation of a cluster's internal rotation axis in single-population and multiple-population clusters). Hereafter, we will more closely investigate the Delta45 model as a reference case. 

Fig. \ref{fig:Theta45_angles} shows the time evolution of the total internal angular momentum orientation for FG and SG stars within the tidal radius of the Delta45 model. The stars in the cluster interact with the tidal field, causing the total internal angular momentum of the system to align with the orbital angular momentum by the end of the simulation. After a small initial change in the direction of the FG's total internal angular momentum, likely due to the effect of the tidal field on the more extended FG population, the two populations are characterised by similar orientations and evolution towards alignment. Figures \ref{fig:ang_sims} and \ref{fig:Theta45_angles} also show small nutations in the total internal angular momentum (see e.g. \citealt{maria18}).

\begin{figure}
	\includegraphics[width=\columnwidth]{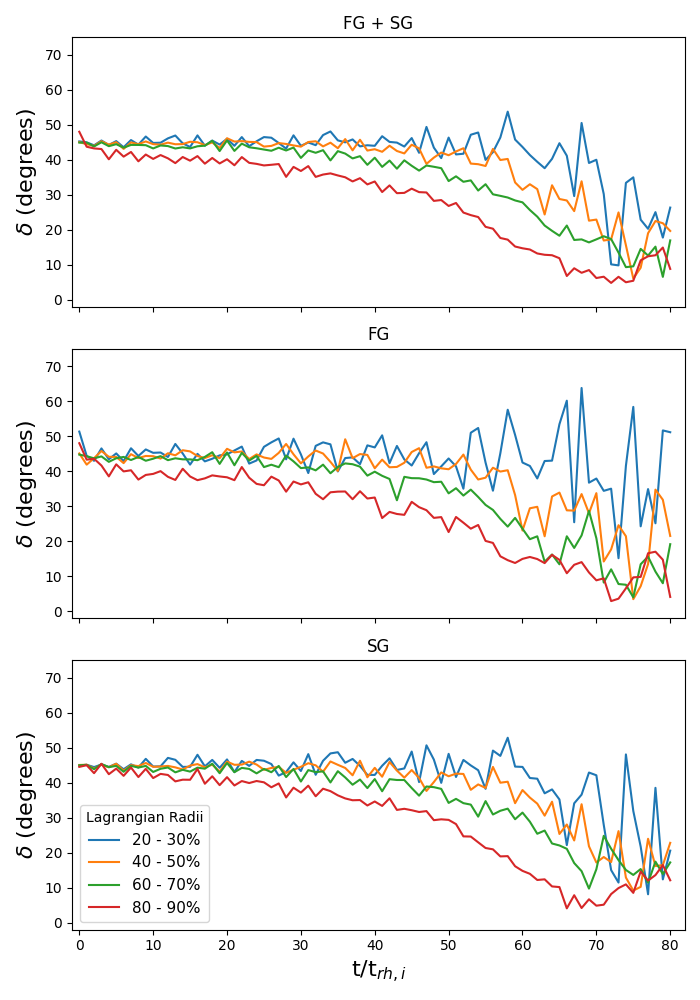}
    \caption{The time evolution of the orientation of the total internal angular momentum vectors for different radial regions is shown for the different subsystems in the Delta45 model. Time is normalised by the initial half-mass relaxation time. Radial regions are set by the Lagrangian radii of all stars (FG + SG), regardless of subsystem. In each subsystem, the total internal angular momentum of stars in the outer regions begins to align with the orbital angular momentum faster than stars in more central regions. }
    \label{fig:ang_regions}
\end{figure}

The effect of the tidal field can be seen more clearly in Fig. \ref{fig:ang_regions}, which shows the evolution of total internal angular momentum orientation for different radial regions in the Delta45 model. In all cases (FG, SG, and FG + SG), the stars in the outer regions begin to align their internal angular momentum with the orbital angular momentum earlier than stars in more central areas of the cluster. The same behaviour is present in both the Delta45 and Delta90 models.

As the system evolves and the two populations dynamically mix, the difference between the direction of their total internal angular momentum is also erased. Overall, the differences between the orientation of the FG and the SG systems are always small and likely difficult to detect observationally. Our results are consistent with those of \cite{Maria} who also found that the total internal angular momentum of the FG and SG initially evolve at different rates, but eventually align with the orbital angular momentum and each other.

\begin{figure}
	\includegraphics[width=\columnwidth]{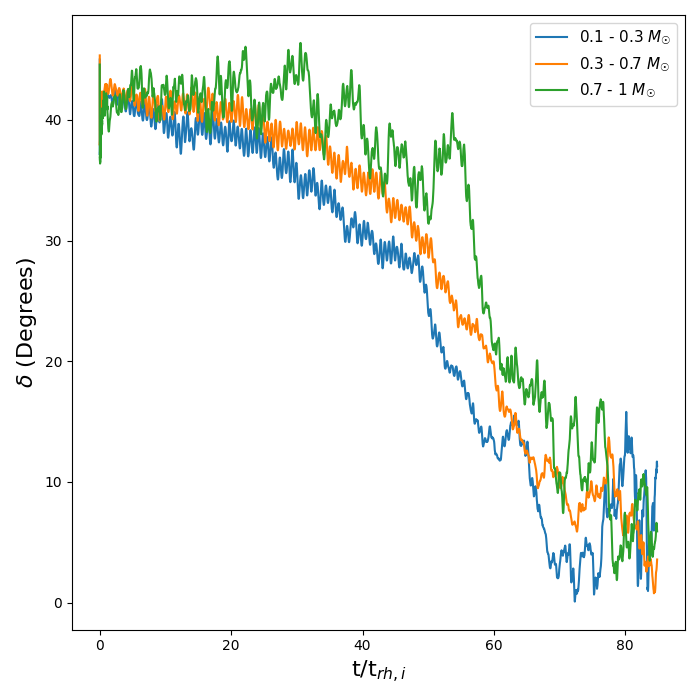}
    \caption{The time evolution of the orientation of the total internal angular momentum vectors for all stars (FG + SG) in the Delta45 model is shown for different mass bins. Time is normalised to the initial half-mass relaxation time. Low-mass stars align with the orbital angular momentum more rapidly than higher mass stars.}
    \label{fig:Theta45_angles_mass}
\end{figure}

We also explore the angular momentum of stars of different masses. Fig. \ref{fig:Theta45_angles_mass} shows the orientation of the total internal angular momentum for stars in different mass groups within the tidal radius of the Delta45 model. Low-mass stars begin to change their $\delta$ faster than higher mass stars. We interpret this as the result of the preferential migration of low-mass stars towards the outer regions of the cluster, where the tidal field can more easily affect their orbital properties. 

\begin{figure}
	\includegraphics[width=\columnwidth]{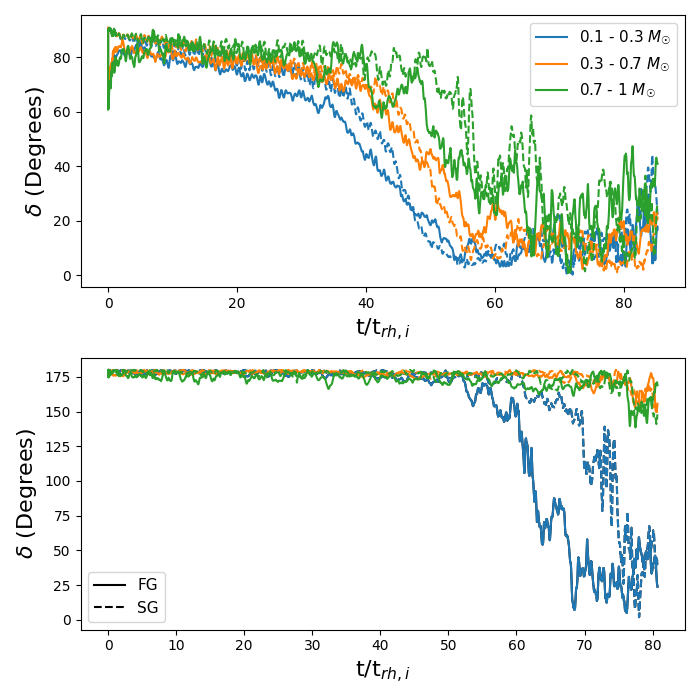}
    \caption{The time evolution of the orientation of the total internal angular momentum vectors for the Delta90 (top) and Delta180 (bottom) models is shown for different mass bins and generations. Time is normalised to the initial half-mass relaxation time. Low-mass stars align with the orbital angular momentum more rapidly than higher mass stars in both models.} 
    \label{fig:ang90_180}
\end{figure}

We find small differences between the direction of internal angular momentum of FG and SG stars in the different stellar mass groups. These differences are present in all cases that do not begin with their internal and orbital angular momentum vectors aligned, but the differences are strongest in the Delta90 and Delta180 models. Fig. \ref{fig:ang90_180} displays the orientation of the total internal angular momentum for different stellar mass groups within the tidal radius of the Delta90 and Delta180 models. There is a small, but consistent difference between the direction of total internal angular momentum of the two generations in the Delta90 model for most of the system's evolution. Delta180 shows the largest difference in the direction of total internal angular momentum between the generations; however, this is only for low-mass stars and only at later dynamical times. Similarly to the behaviour seen in Fig. \ref{fig:Theta45_angles_mass}, Fig. \ref{fig:ang90_180} also shows that low-mass stars begin to align with the direction of the orbital angular momentum faster than stars of higher masses.

It is interesting to note that, in each simulation, the generations are spatially mixed by the time the model's direction of total internal angular momentum undergoes significant change  (\textasciitilde40 $t_{\rm rh,i}$). 
This suggests that the difference in rotation velocities between generations may be a driving factor behind the delay in alignment of the different generations.

The SG's higher rotation velocity may cause the SG to take a longer time to align with the direction of orbital angular momentum. We stress again that while for many of the N-body models presented in this study there is a difference between the direction of total internal angular momentum of the two generations, the difference is small and would likely be difficult to detect observationally.

\begin{figure}
	\includegraphics[width=\columnwidth]{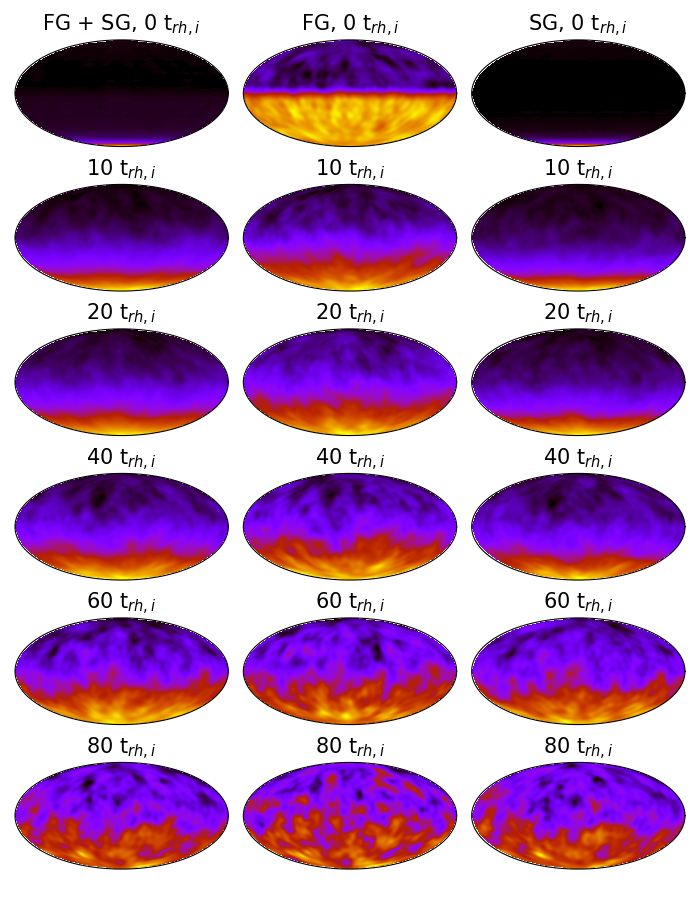}
    \caption{Internal angular momentum maps are shown for FG + SG (left), FG (middle), and SG (right) for the Delta0 model. 
    The colour map corresponds to the concentration of stars whose internal angular momentum points in a given direction. Higher concentrations are represented by a yellow colouration, while lower concentrations are represented by a black colouration. A tophat smoothing has been applied to the density maps. Rows show representative time steps throughout the simulation. The orbital angular momentum vector is aligned with the bottom of the maps.}
    \label{fig:ang_map0}
\end{figure}

\begin{figure}
	\includegraphics[width=\columnwidth]{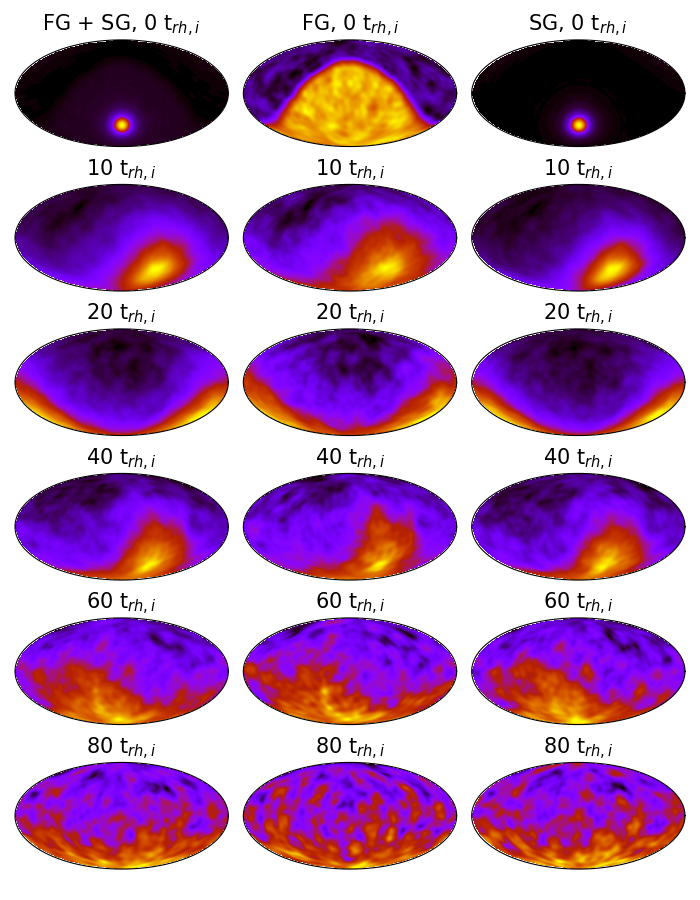}
    \caption{Similar to Fig. \ref{fig:ang_map0} but for the Delta45 model.}
    \label{fig:ang_map45}
\end{figure}

\begin{figure}
	\includegraphics[width=\columnwidth]{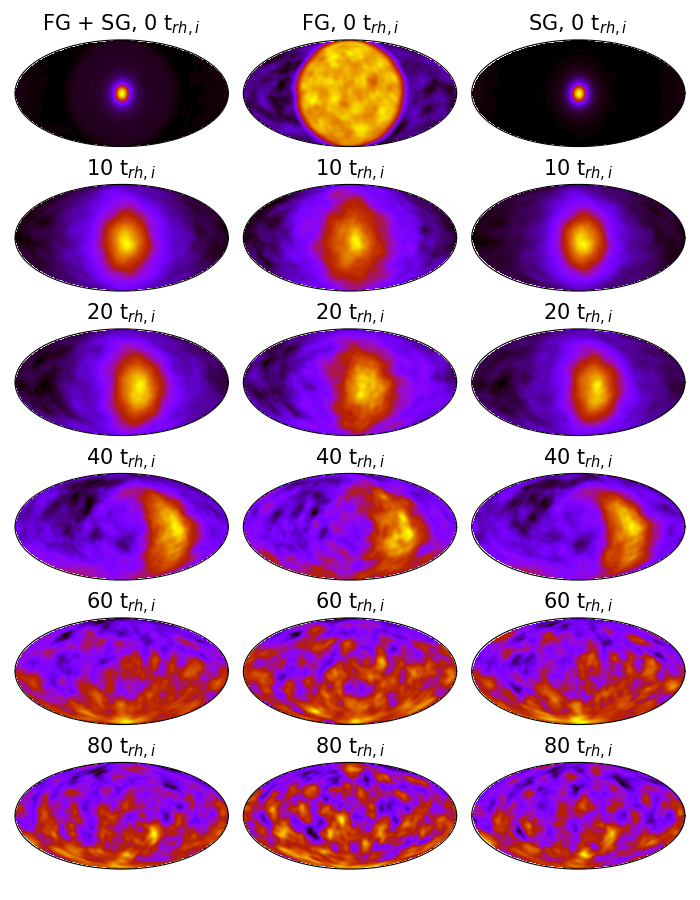}
    \caption{Similar to Fig. \ref{fig:ang_map0} but for the Delta90 model}
    \label{fig:ang_map90}
\end{figure}

\begin{figure}
	\includegraphics[width=\columnwidth]{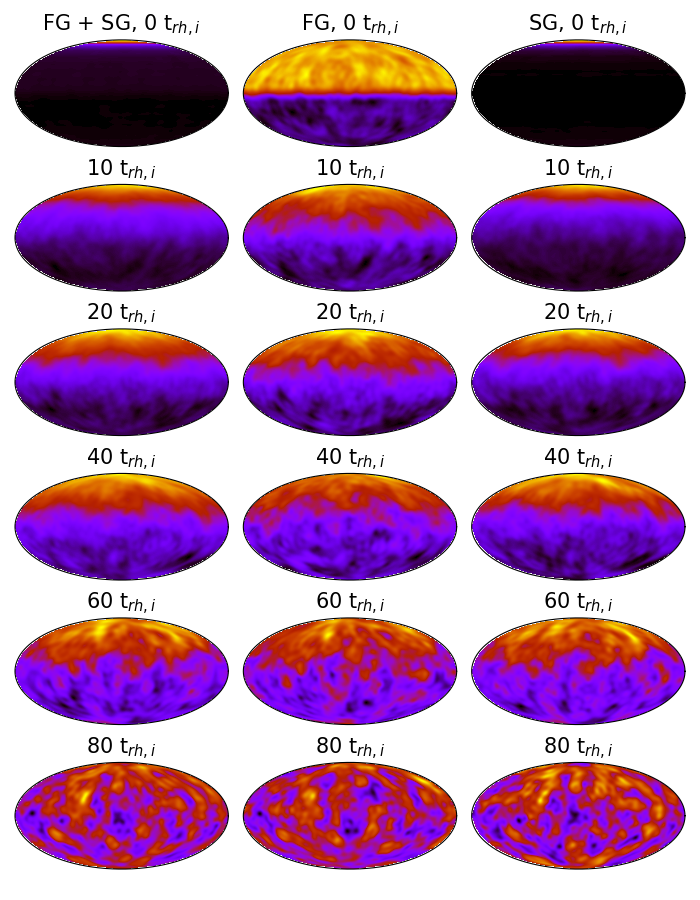}
    \caption{Similar to Fig. \ref{fig:ang_map0} but for the Delta180 model. Unlike the other models, the peak direction of internal angular momentum does not fully align with the direction of the orbital angular momentum by the end of the simulation for the FG, SG, or FG + SG.}
    \label{fig:ang_map180}
\end{figure}

The results of the analysis presented thus far focus on the total internal angular momentum of the stars in the models. In order to gain further insight into the models' kinematics, we examine the distribution of the orientation of internal angular momentum for the individual stars. Figures \ref{fig:ang_map0}, \ref{fig:ang_map45}, \ref{fig:ang_map90}, and \ref{fig:ang_map180} utilise a method described in \cite{IT21} to create internal angular momentum maps of stars of different generations. 
Each map shows the distribution of the orientation of the individual stars' internal angular momentum at the given time.
The maps present a 2D projection of the histogram of the internal angular momentum distribution with brighter yellow areas corresponding to angles of angular momentum orientation with a larger fraction of stars. The orbital angular momentum vector is aligned with the bottom of the maps.

The centre of the map is aligned with the direction of the galactic centre. 
The first row of these figures demonstrates that the density of the direction of internal angular momentum when considering all stars is initially dictated by the highly rotating SG. The density of the direction of the internal angular momentum becomes more diffuse as the system evolves and the generations become spatially and kinematically mixed. The density of orientations for the two generations appears nearly identical after approximately 40 $t_{\rm rh,i}$.

Fig. \ref{fig:ang_map0} shows the difference in the spread of orientation of the internal angular momentum between generations. Nonetheless, the location of the peak in the distribution of the orientations stays constant due to the internal angular momentum already being aligned with the direction of the orbital angular momentum. 
Figures \ref{fig:ang_map45}, \ref{fig:ang_map90}, and \ref{fig:ang_map180} show the evolution of the distribution of the orientation of the internal angular momentum within the system as well as the difference between the distribution of the internal angular momentum of the two generations. Each model shows that there is an initially large difference in the distribution of the orientation of internal angular momentum between the stars of the two generations. The disk-like, high-rotation structure of the SG causes the distribution of orientations to be more peaked around the initial angular momentum orientation, while the more diffuse, lower rotation FG has a larger spread in the direction of internal angular momentum for stars within the generation. That difference in the density of orientations is gradually erased as the FG and SG subsystems mix. 
The Delta45 and Delta90 models clearly display the effects of the tidal field on the orientation of internal rotation as the peak density and overall distribution of internal angular momentum orientation moves to align with the direction of the orbital angular momentum at the bottom of the maps. In this way, these maps present another view of what is shown in Fig. \ref{fig:ang_sims}.

While the Delta0, Delta45, and Delta90 models all end with the orientation of their internal angular momentum aligned with the orbital angular momentum, the same is not true of the Delta180 model. At the end of the simulation, the orientation of the internal angular momentum is still counter-aligned with the direction of the orbital angular momentum. The final time-step of Fig. \ref{fig:ang_map180}, however, shows that there is some diffusion away from this orientation. High density regions on these maps at 80 $t_{\rm rh,i}$ show that stars have begun to change their orientations and move towards aligning with the direction of orbital angular momentum. 

This connects to what is shown in Fig. \ref{fig:ang90_180} where the total internal angular momentum vector of low-mass stars within the tidal radius has largely aligned with the orbital angular momentum. While many of the stars in the Delta180 model still maintain an internal angular momentum direction which is counter-aligned with the direction of the orbital angular momentum, low-mass stars largely have the direction of their internal angular momentum aligned with the orbital angular momentum by the end of the simulation. 

\begin{figure}
	\includegraphics[width=\columnwidth]{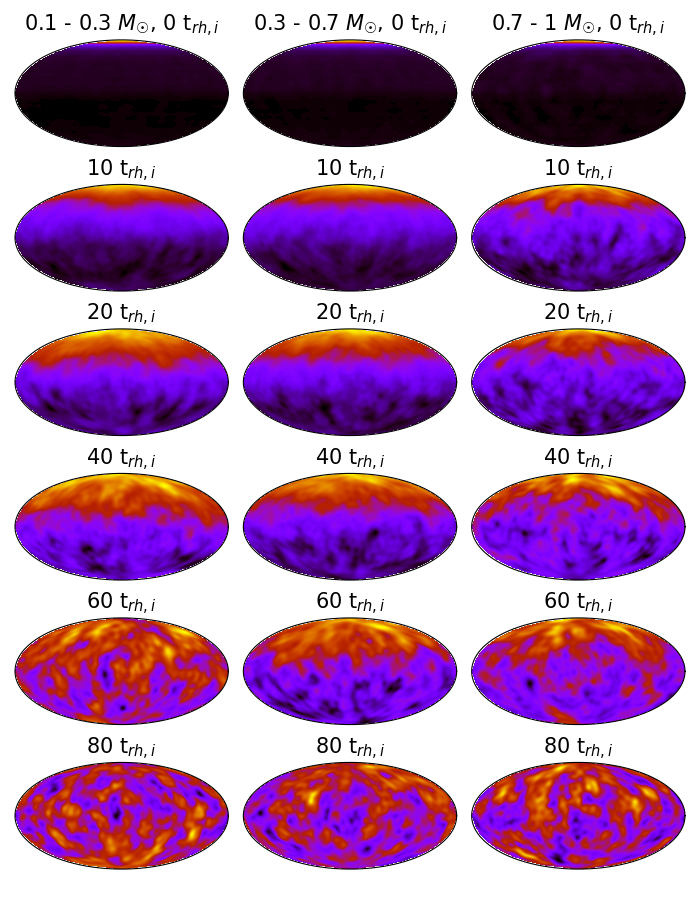}
    \caption{Similar to Fig. \ref{fig:ang_map180} but for individual mass bins of all stars (FG + SG) in the Delta180 model. The density of the orientation of internal angular momentum for low-mass stars is much closer to aligning with the direction of the orbital angular momentum than that of the intermediate- or high-mass stars.}
    \label{fig:ang_map180_mass}
\end{figure}

We can further explore this in Fig. \ref{fig:ang_map180_mass}. Here we investigate the distribution of the direction of internal angular momentum for different mass bins in the Delta180 model. It can clearly be seen that the low-mass stars are aligned more closely with the direction of the orbital angular momentum, while the intermediate and high-mass stars maintain an orientation similar to that of the initial conditions.

\begin{figure}
	\includegraphics[width=\columnwidth]{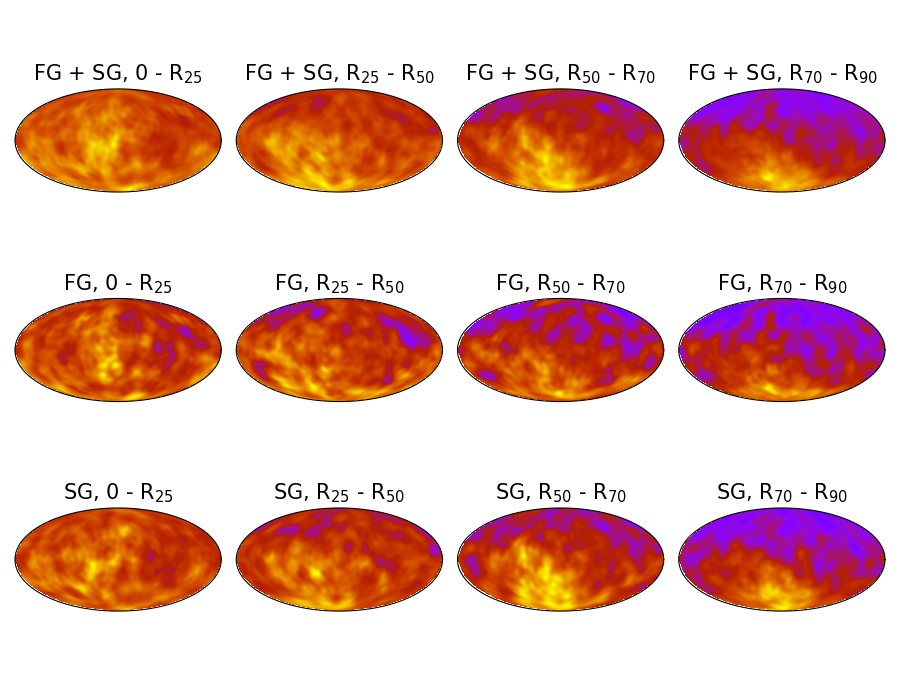}
    \caption{Angular momentum maps are shown for FG + SG (top), FG (middle), and SG (bottom) for different radial regions in the Delta45 model at 60 $t_{\rm rh,i}$. The internal angular momentum of stars in the outer regions are shown to align more closely with the direction of the orbital angular momentum.}
    \label{fig:ang_map_region}
\end{figure}

Finally, we utilise these maps to show an alternative view of what is displayed in Fig. \ref{fig:ang_regions}. Fig. \ref{fig:ang_map_region} shows that the peak density of the orientation of internal angular momentum, as well as the overall distribution of orientations, becomes increasingly aligned with the direction of the orbital angular momentum towards the outer regions of the cluster over time. Again, this is due to the effects of the tidal field interactions being stronger towards the outer regions of the system.

\section{Conclusions}
In this paper we have investigated the dynamical evolution of rotating multiple-population globular clusters. Our study is based on a suite of N-body simulations and addresses a number of questions concerning the long-term evolution of the differences between the dynamical properties of first- and second-generation stars imprinted at the time of their formation. Informed by a number of studies of the formation of multiple stellar populations in rotating clusters (see e.g. \citealt{bekki10}, \citeyear{bekki11}; \citealt{lacchin22}), our simulations start with clusters characterised by a rapidly rotating and flattened second-generation subsystem concentrated in the inner regions of a slowly-rotating, spherical first-generation cluster. In order to investigate the interplay between the effects of the external tidal field on the clusters' internal rotation, we have studied the evolution of systems starting with different initial inclinations of the internal angular momentum relative to the cluster's orbital plane.

The results of our study are as follows. 

\begin{itemize}
    \item We explore the models' dynamical evolution through the evolution of their mass density, rotation velocity, and velocity anisotropy profiles. By examining the mass density profiles, we see that the spatial differences between generations are greatly diminished within a few initial half-mass relaxation times (Fig. \ref{fig:masden_prof}). We find that, even after several initial half-mass relaxation times, there is still a difference in the rotation velocities of the two generations (Fig. \ref{fig:rot_prof}). However, it is important to emphasise that initial differences between the rotation of FG and SG stars rapidly decrease and, even in dynamically young clusters with ages equal to just a few half-mass relaxation times, these differences may be significantly smaller than those imprinted at the time of the clusters' formation. This is the case for all models (Fig. \ref{fig:max_rot_sims}) and for stars with different masses (Figures \ref{fig:rot_max_dif_mass} $\&$ \ref{fig:max_rot_sims_mass}). This suggests that caution is necessary in drawing conclusions concerning the primordial rotation differences and their extent based on present-day kinematics, even when considering relatively dynamically young clusters (see also \citealt{Maria}; \citealt{dalessandro24} for further discussion of this point.) 
   
    In agreement with previous investigations, our analysis also reveals differences between the anisotropy of the velocity distribution of the FG and SG populations. SG stars are characterised by a stronger velocity anisotropy than that of FG stars. Although these differences also decrease during the clusters' long-term evolution, they tend to persist longer than those between FG and SG rotational kinematics. Observational studies with wide fields spanning a broad range of clustercentric differences are more likely to detect differences between the FG and SG spatial and kinematic differences. Studies limited to the inner regions may fail to reveal dynamical differences or may reveal much smaller differences than those present in the intermediate and outer regions. 
    \item By studying the time evolution of the rotation curves for different stellar masses and the resulting peak velocities, we find a dependence of rotation velocity on stellar mass. Higher-mass stars maintain a higher peak rotation velocity than low-mass stars. We also see that the kinematic differences between generations are erased more efficiently for high-mass stars as compared to low-mass stars (see Figures \ref{fig:rot_prof_mass} and \ref{fig:rot_max_dif_mass}). We therefore find that the inclusion of low-mass stars would improve the chances of detecting kinematic differences observationally.
    
    \item By exploring the interaction of the models' internal rotation with the galactic tidal field, we find that models whose internal rotation is not initially aligned or counter-aligned with the direction of their orbital angular momentum may maintain differences between the rotational kinematics of FG and SG stars for a longer time (Fig. \ref{fig:max_rot_sims}). These differences between generations are present predominantly in low-mass stars (Fig. \ref{fig:max_rot_sims_mass}). Our analysis reveals, however, that these differences may be small and difficult to detect observationally. The observational detection may be hindered by various factors such as, for example, errors in velocity measurements, limited statistical samples, possible misclassification of FG and SG stars (see also \citealt{maria18} for possible problems due to projection effects and observations of clusters along different lines of sight).
   
    \item We explore the interaction of the model's internal angular momentum and the galactic tidal field. We find that, in agreement with previous studies, the tidal field can cause the internal angular momentum of globular clusters to align with the direction of their orbital angular momentum over time. Models Delta45 and Delta90 largely align their internal angular momentum with the direction of their orbital angular momentum by the end of our simulations. The direction of internal angular momentum for Delta0, which begins with its internal and orbital angular momentum aligned, shows little variation throughout the simulation (see Figures \ref{fig:ang_sims}, \ref{fig:Theta45_angles}, \ref{fig:ang_map0}, \ref{fig:ang_map45}, and \ref{fig:ang_map90}). The direction of internal angular momentum for the Delta180 model, however, remains largely counter-aligned with the direction of the orbital angular momentum even at the end of the simulation (Fig. \ref{fig:ang_map180}). 
    \item We also explore the mass dependence of the interaction between the cluster's internal angular momentum and the galactic tidal field. In our analysis, we determine that low-mass stars align their internal angular momentum with the direction of the cluster's orbital angular momentum more rapidly than higher-mass stars. In addition, while the internal angular momentum of the Delta180 model does not align with the direction of the orbital angular momentum by the end of the simulation when considering stars of all masses, the internal angular momentum of the low-mass stars of that model does (see Figures \ref{fig:Theta45_angles_mass}, \ref{fig:ang90_180}, and \ref{fig:ang_map180_mass}). We interpret this as the result of the preferential migration of low-mass stars towards the model's outer regions where the effects of the tidal field are stronger. 
     
    \item By exploring the orientation of the models' internal angular momentum in different radial regions, we find - in agreement with previous studies - that there is a radial dependence of the direction of internal angular momentum within a cluster. The galactic tidal field more strongly affects stars at larger radii, leading those stars to align the direction of their internal angular momentum with the direction of the orbital angular momentum faster than stars closer to the centre of the cluster (Figures \ref{fig:ang_regions} and \ref{fig:ang_map_region}).

\end{itemize}

\begin{acknowledgements}
      EV acknowledges support from NSF grant AST-2009193. EV acknowledges also support from the John and A-Lan Reynolds Faculty Research Fund. ED acknowledges financial support from the INAF Data Analysis Research Grant (ref. E. Dalessandro) of the "Bando Astrofisica Fondamentale 2024". ALV acknowledges support from a UKRI Future Leaders Fellowship (MR/X011097/1). We thank the referee for a thorough review of this paper. 
\end{acknowledgements}

\bibliographystyle{aa}
\bibliography{example}

\end{document}